\documentclass[aps,pre,twocolumn,preprint,superscriptaddress]{revtex4-1}
\usepackage{amsfonts}

\usepackage{eurosym}
\usepackage[utf8]{inputenc}
\usepackage{amsmath}
\usepackage{amssymb}
\usepackage{latexsym}
\usepackage{ifpdf}
\usepackage{esint}
\usepackage{dsfont}
\usepackage{slashed}
\usepackage{color}
\usepackage[normalem]{ulem}

\begin{document}

\title{Reply to the Comment on ``The Lifshitz-Matsubara sum formula for the
Casimir pressure between magnetic metallic mirrors''}
\author{R. Gu\'erout}
\affiliation{Laboratoire Kastler Brossel, UPMC-Sorbonne Universités, CNRS, ENS-PSL Research University, 
Collège de France, F-75252 Paris, France}
\email{guerout@lkb.upmc.fr}
\author{A. Lambrecht}
\affiliation{Laboratoire Kastler Brossel, UPMC-Sorbonne Universités, CNRS, ENS-PSL Research University, 
Collège de France, F-75252 Paris, France}
\author{K. A. Milton}
\affiliation{H. L. Dodge Dept. of Physics and
Astronomy, Univ. of Oklahoma, Norman, OK 73019 USA}
\author{S. Reynaud}
\affiliation{Laboratoire Kastler Brossel, UPMC-Sorbonne Universités, CNRS, ENS-PSL Research University, 
Collège de France, F-75252 Paris, France}
\date{\today }

\begin{abstract}
We reply to the "Comment on `The Lifshitz-Matsubara sum formula for the Casimir pressure between magnetic metallic mirrors'". We
believe the comment misrepresents our papers, and fails to provide a plausible resolution to the conflict between theory and
experiment.
\end{abstract}

\maketitle

In our recent publication~\cite{Guerout2016}, we have extended the analysis 
presented in~\cite{Guerout2014} for non magnetic materials in order
to include the case of magnetic materials. 
The Comment~\cite{Klimchitskaya2016} on the paper~\cite{Guerout2016} 
does not discuss this extension to the magnetic case.
It deals with points related to the context in which our 
paper~\cite{Guerout2014} was written but misses completely the original 
results obtained in~\cite{Guerout2014} as well as~\cite{Guerout2016}. 
It also contains descriptions of the content of~\cite{Guerout2016,Guerout2014} 
which clearly contradicts what is written in these papers. 

The authors of~\cite{Klimchitskaya2016} claim that we advocated a redefinition 
of the plasma susceptibility. In contrast to this claim, we were extremely careful 
to avoid confusion between different definitions of the susceptibility. To this purpose,
we introduced specific notations in order to distinguish the 3 different definitions
\begin{itemize}
\item $\chi_\gamma=\epsilon_\gamma-1=\frac{\sigma_\gamma}{-i\omega}$ with 
$\sigma_\gamma=\frac{\omega^2_p}{\gamma-i\omega}$ for the Drude model~; 
\item $\chi_\eta$ for the limit of the Drude model when $\gamma\to0$
\item $\chi_0$ for the lossless plasma model where $\gamma=0$. 
\end{itemize}

We clearly stated in~\cite{Guerout2016,Guerout2014} that the Drude model 
matches at low frequencies the optical (permittivity $\epsilon_\gamma$) 
and electrical (conductivity $\sigma_\gamma$) 
characteristic functions of the metallic plates used in the experiment when 
$\gamma$ has the appropriate value for the metal of interest (say for example gold or nickel).  
In contrast, the models $\chi_0$ as well as $\chi_\eta$ do not match these 
optical and electrical properties~\cite{Lambrecht2000,Svetovoy2008}. 
We introduced the words ``Casimir puzzle'' to emphasize the undisputed fact 
that experimental measurements~\cite{Decca2007,Chang2012} are in better
agreement with the plasma model $\chi_0$ than with the Drude model $\chi_\gamma$
whereas the latter is a much better motivated description of the actual properties of the plates.

The main content of our papers~\cite{Guerout2016,Guerout2014} consists in careful 
derivations of the Lifshitz formulas for the Casimir force when these different definitions are used. 
In particular, we devoted a special attention to give a proper description of the
difference between $\chi_\eta$ and $\chi_0$, in the sense of distribution theory.
We did then show that this difference is a source of delicate problems
in the usual derivation of the Casimir pressure, and that the commonly used
Lifshitz-Matsubara sum formula has to be corrected when using the
susceptibility $\chi_\eta$. This is the main technical result in our papers.
The Comment~\cite{Klimchitskaya2016} gives an unfair representation of their
content by claiming that we confuse the 2 definitions $\chi_\eta$ and $\chi_0$
and then ignoring the new results obtained through a careful study of their difference. 

Our papers are interesting from a pure theoretical point of view, as they shed new light 
on the subtleties of the application of Cauchy's residue theorem in the context of the
calculation of the Casimir forces. In contrast the Comment contains confusing discussions 
of analytical properties of response functions, based on textbook material (ref.~[10] 
in~\cite{Klimchitskaya2016}) which is not sufficient for a correct analysis of the debated points. 
Equation (4) in~\cite{Klimchitskaya2016} is not a sensible relation in the theory of distributions,
and its loose manipulation leads to the difficulties discussed in the Comment.
In contrast, our definition of $\chi_\eta$ (eq.~(8) in~\cite{Guerout2016}) or its representation
 in terms of the distribution $\delta^\prime$ (eq.~(9) in~\cite{Guerout2016} fully equivalent to 
eq.~(8)) allows one to derive the correct formula for the Casimir force~\cite{Guerout2014}.

The last part of the Comment is an awkward justification of the use of the ill-motivated
plasma model when comparing experimental results to theoretical predictions.
The susceptibility function (6) in~\cite{Klimchitskaya2016} does not match the 
well-known optical and electrical properties of metallic plates, and the agreement 
of the predictions drawn from this model with experiments can only be considered 
as a puzzle yet to be solved. The proposition in~\cite{Klimchitskaya2016} according to which
metals would not respond to quantum fluctuations of fields as they do for classical fields
is weird as it invalidates the theory used to obtain predictions.
The very idea of comparing theory and experiment loses any meaning if one feels free 
to change at will the theory so that its predictions agree with the measurements.


\end{document}